\documentclass[aps,twocolumn,pra,floatfix,amsmath,amssymb,superscriptaddress,showpacs]{revtex4-1}
\usepackage{graphicx}
\usepackage{dcolumn}
\usepackage{epstopdf}
\usepackage{bm}
\usepackage{color}
\usepackage{ulem}

\def\jcp#1#2#3{J.~Chem.~Phys.~{\bf #1},\ #2\ (#3)}

\def\pr#1#2#3{Phys.~Rev.~{\bf #1},\ #2\ (#3)}
\def\pra#1#2#3{Phys.~Rev.~A~{\bf #1},\ #2\ (#3)}
\def\prb#1#2#3{Phys.~Rev.~B~{\bf #1},\ #2\ (#3)}

\def\pre#1#2#3{Phys.~Rev.~E~{\bf #1},\ #2\ (#3)}
\def\rmp#1#2#3{Rev.~Mod.~Phys.~{\bf #1},\ #2\ (#3)}
\def\prl#1#2#3{Phys.~Rev.~Lett.~{\bf #1},\ #2\ (#3)}

\def\nat#1#2#3{Nature~{\bf #1},\ #2\ (#3)}

\def\njp#1#2#3{New~J.~Phys.~{\bf #1},\ #2\ (#3)}
\newcommand{\etal}{{\it et al.}}

\begin{document}
\title{Transport of quantum excitations via local and nonlocal fluctuations}

\author{M. Zhang}
\affiliation{ITAMP, Harvard-Smithsonian Center for Astrophysics, Cambridge, Massachusetts 02138, USA}
\affiliation{Department of Physics, Beijing Normal University, Beijing, 100875, P. R. China}
\author{Tony E. Lee}
\affiliation{ITAMP, Harvard-Smithsonian Center for Astrophysics, Cambridge, Massachusetts 02138, USA}
\affiliation{Department of Physics, Harvard University, Cambridge, Massachusetts 02138, USA}
\author{H. R. Sadeghpour}
\affiliation{ITAMP, Harvard-Smithsonian Center for Astrophysics, Cambridge, Massachusetts 02138, USA}

\date{\today}

\begin{abstract}
In quantum systems, one usually seeks to minimize dephasing noise and disorder. The efficiency of transport in a quantum system is usually degraded by the presence of noise and disorder. However, it has been shown that the combination of the two can lead to significantly more efficiency than each by itself. Here, we consider how the addition of nonlocal noise, in the form of incoherent hopping, affects the transport efficiency. We show that incoherent hopping introduces additional local extrema in the efficiency function and investigate the crossover from a quantum random walk to a classical random walk.
\end{abstract}

\maketitle

\section{Introduction}
Environmental noise is simultaneously intrinsic to any physical process and usually detrimental to its efficient operation. In the context of excitation transport in quantum systems, dephasing noise leads to decoherence and reduces the transport efficiency. However, in systems possessing spatial or dynamical symmetry breaking, fluctuating noise can at times enhance transport, as in ratchets and Brownian motors~\cite{hanggi_augsburg_2009_rmp_81_387}. The idea of stochastic resonance has been used to describe transport and power enhancement in a variety of classical, biological and quantum networks~\cite{haenggi}. Among those, a natural light-harvesting system exhibiting extremely high energy transfer efficiency has attracted a lot of attentions and even helped precipitate the field of quantum biology~\cite{ball}. The mechanism of excitation energy transfer (EET) in light harvesting complexes and to what extent EET is influenced by quantum mechanical effects, calls for a consistent theoretical model~\cite{fleming}.

Recently, it has been pointed out that for realistic coupling strength in photosynthetic pigments, quantum and classical coherent transports are identical~\cite{eisfeld11pre}. Previously, a quantum walk~\cite{lloyd} framework was already presented for disordered systems, in which under the influence of quantum interference, an interplay between free Hamiltonian evolution and thermal fluctuation in the environment leads to an increase in EET. In the absence of noise, there is hardly any transport since disorder leads to Anderson localization of eigenstates. However, the addition of a local dephasing noise scrambles the wave function so that it is no longer localized in space, leading to increased transport efficiency~\cite{lloyd09}. In Ref.~\cite{plenio}, the dephasing-assisted transport in quantum networks and biomolecules was investigated assuming a similar system-environment interaction. They showed that, reminiscent to stochastic resonance in classical system~\cite{wiesenfeld}, transport of excitations across dissipative quantum networks can be enhanced by local dephasing noise.

In this work, we study the effect of nonlocal noise, in the form of incoherent hopping, on the transport efficiency in a prototypical one dimensional spin chain. We obtain analytical and numerical results for transport efficiency in quantum chains of varying sizes. We show that in the absence of disorder, the addition of incoherent hopping leads to a local minimum in the efficiency, marking the crossover from coherent quantum to incoherent classical dynamics. We show that in a disordered chain, incoherent hopping leads to a local minimum and a local maximum. We also explore an experimental implementation of such noise-enhanced transport dynamics with trapped ions.

\section{General Formalism}\label{sec_model}
We adopt the quantum walk framework of Refs.~\cite{plenio,lloyd} to a one-dimensional chain of $N$ sites subject to dissipation. Each site is a spin-1/2 with an energy difference $\omega_{k}$ between the two spin states. The Hamiltonian and the Lindblad superoperators are
\begin{eqnarray}
H &=& \sum_{k=1}^N\omega_{k}\sigma_{k}^{+}\sigma_{k}^{-}+v\sum_{\langle k\ell\rangle}\sigma_{k}^{+}\sigma_{\ell}^{-},
\end{eqnarray}
\begin{eqnarray}
\mathcal{L}_{\text{diss}}(\rho) &=& \Gamma\sum_{k=1}^{N}\left[-\frac{1}{2}\left\{\sigma_{k}^{+}\sigma_{k}^{-},\rho\right\}+\sigma_{k}^{-}\rho\sigma_{k}^{+}\right],
\end{eqnarray}
\begin{eqnarray}
\mathcal{L}_{\text{deph}}(\rho) &=& \gamma\sum_{k=1}^{N}\left[-\frac{1}{2}\left\{\sigma_{k}^{+}\sigma_{k}^{-}\sigma_{k}^{+}\sigma_{k}^{-},\rho\right\}+\sigma_{k}^{+}\sigma_{k}^{-}\rho\sigma_{k}^{+}\sigma_{k}^{-}\right]\nonumber\\
&&+ \gamma_{h}\sum_{\langle k\ell\rangle}\Big[-\frac{1}{2}\left\{\sigma_{k}^{+}\sigma_{\ell}^{-}\sigma_{\ell}^{+}\sigma_{k}^{-},\rho\right\}\nonumber\\ &&\quad\quad\quad\quad+\sigma_{\ell}^{+}\sigma_{k}^{-}\rho\sigma_{k}^{+}\sigma_{\ell}^{-}\Big],
\end{eqnarray}
where ${\langle k\ell\rangle}$ denotes summing over nearest neighbors and $\{A,B\}=AB+BA$. There is coherent hopping $v$ between nearest neighbors. Each site dissipates energy into the environment at a rate $\Gamma$ when it is in the upper energy state, corresponding to the decay of an excitation. In addition to the local dephasing rate $\gamma$ already considered in previous studies~\cite{lloyd,plenio}, we also include an incoherent hopping rate $\gamma_h$. This rate is so-called since it originates from the fluctuations in the nearest-neighbor hopping. We assume white-noise properties in these exterior field so that the system undergoes a Markov evolution as in Ref.~\cite{yuting}.

To study the energy transfer, we add an additional site $N+1$, which is dissipatively coupled to site $N$ with a rate $\Lambda$,
\begin{eqnarray}
\mathcal{L}_{\text{sink}}(\rho) &=& \Lambda\Big[-\frac{1}{2}\left\{\sigma_{N}^{+}\sigma_{N+1}^{-}\sigma_{N+1}^{+}\sigma_{N}^{-},\rho\right\}\nonumber\\ &&\quad\quad + \sigma_{N+1}^{+}\sigma_{N}^{-}\rho\sigma_{N}^{+}\sigma_{N+1}^{-}\Big].
\end{eqnarray}
Site $N+1$ is a sink as once the excitation enters, it is removed from the system.

The Lindblad master equation for the density matrix $\rho$ reads
\begin{equation}
\frac{\mathrm{d}\rho}{\mathrm{d}t}=-i\left[H, \rho\right]+\mathcal{L}_{\text{diss}}(\rho)+\mathcal{L}_{\text{deph}}(\rho)+\mathcal{L}_{\text{sink}}(\rho),
\end{equation}
This is basically a set of $N^{2}$ linear ordinary differential equations with initial conditions $\rho_{0}$ and thus can be conveniently solved with the Laplace transformation.

We assume that the chain is initially excited at site $1$. The total population depends only on $\Gamma$ and $\Lambda$, while the relative population between sites depends on the other parameters. Throughout this work, we choose $\Lambda/\Gamma=10.0$ and $\Gamma/v=0.02$.

Transfer efficiency is defined as the probability of excitations arriving at the sink,
\begin{equation}
\eta = \Lambda\int_0^{\infty}\rho_{N,N}(\tau)\mathrm{d}\tau.
\end{equation}
Below, we explore how the inclusion of incoherent hopping affects the efficiency of transfer.

\section{Uniform system}\label{sec_uniform}
We start with a chain of $N=2$ sites. In a rotating frame, the coherent dynamics of this system is governed by two parameters, the coherent hopping $v$ and the difference in site energies $\delta=\omega_{1}-\omega_{2}$. It is well-known that in a closed quantum system, coherent hopping favors excitation transfer while disorder tends to localize excitations~\cite{anderson}. Therefore, the closed quantum system has two inherent competing energy transfer processes. It has been shown that one can counteract the localizing effect of disorder by including dephasing $\gamma$.

Using Laplace transformation, the EET efficiency can be obtained analytically,
\begin{eqnarray}\label{eq_eta_2site}
\eta&=&\frac{\Lambda}{2\Gamma+\Lambda}\left[1-\frac{\Gamma(\Gamma+\Lambda)}{\Gamma(\Gamma+\Lambda)+(4v^{2}/\Delta+\gamma_{h})(2\Gamma+\Lambda)}\right],\nonumber\\ \\
\Delta&=&\mathcal{D}+\frac{4\delta^{2}}{\mathcal{D}},\\
\mathcal{D}&=&2\Gamma+\Lambda+2(\gamma+\gamma_{h}).
\end{eqnarray}
Here, $\mathcal{D}$ is the sum of all incoherent rates. 

\begin{figure}[b]
\begin{center}
\includegraphics[width=.9\linewidth]{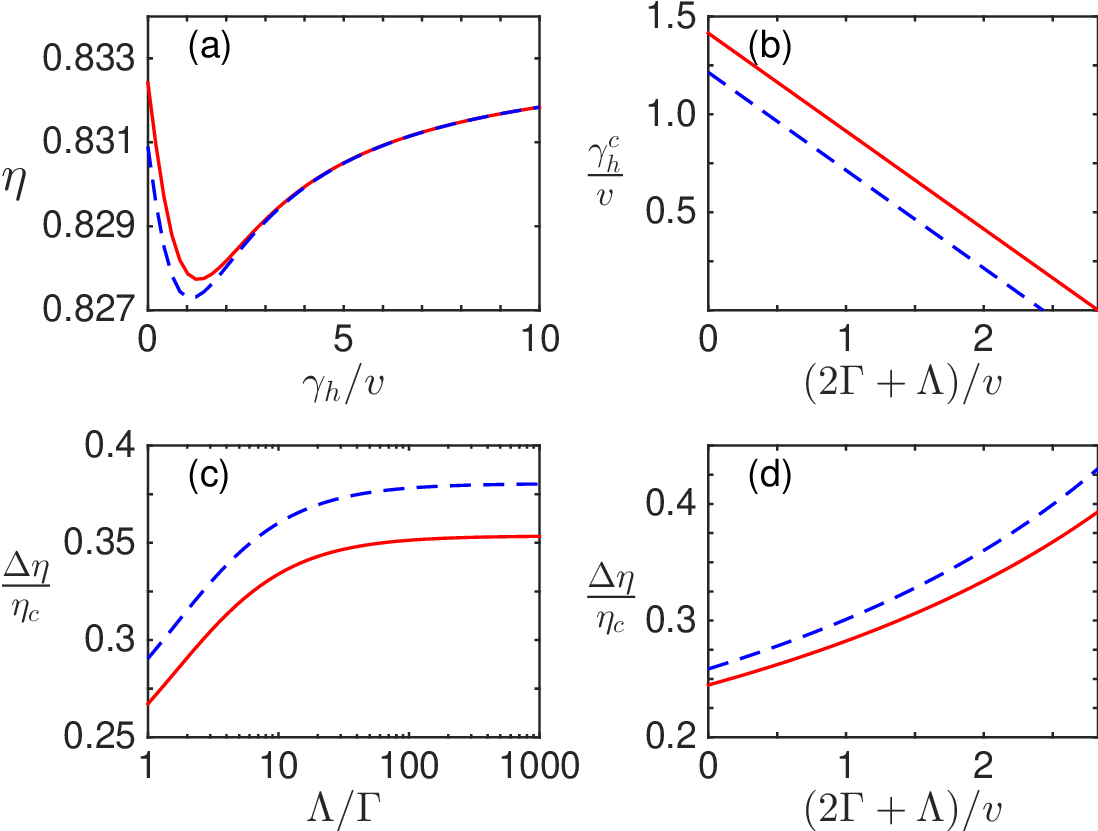}
\caption{(Color online) Competition of coherent and incoherent hopping for a uniform ($\delta=0$) two-site chain. (a) Energy transfer efficiency as a function of ratio $\gamma_{h}/v$ for $\Lambda/\Gamma=10$ and $\Gamma/v=0.02$; The minimum near $\gamma_{h}=v$ marks the transition from a quantum to a classical random walk. (b) Dependence of dip position on total dissipation rate; (c) Dependence of relative dip depth on $\Lambda/\Gamma$ for $\Gamma/v=0.02$; and (d) Dependend of relative dip depth on $(2\Gamma+\Lambda)/v$ for $\Lambda/\Gamma=10$. Red solid lines are for zero local dephasing and blue broken for $\gamma/v=0.2$.}
\label{fig_2site_nodisorder}
\end{center}
\end{figure}

First, we set disorder $\delta=0$ to see the competition between coherent hopping $v$ and incoherent hopping $\gamma_{h}$. Since local dephasing $\gamma$ only shifts the total dissipation, for simplicity, we also set it to be zero. Straightforward algebra shows that for a given set of $\Gamma $ and $\Lambda$, the efficiency as a function of $\gamma_{h}/v$ has a local minimum, as seen in Fig.~\ref{fig_2site_nodisorder}(a). The reason for the minimum is as follows. When $\gamma_{h}$ is zero, the dynamics is fully coherent, and the excitation undergoes a quantum random walk. As $\gamma_{h}$ increases, the efficiency decreases since the incoherent hopping destroys the coherence required for a quantum random walk. However, when $\gamma_{h}$ is sufficiently large, the dynamics is dominated by incoherent hopping and the excitation executes a classical random walk. The efficiency minimum marks the transition from a quantum to a classical random walk.

To characterize the local minimum, we define the depth of the dip as the difference between the minimum $\eta_\textrm{min}$ and the saturated efficiency $\eta_{c}$ when either coherent or incoherent hopping is much stronger than dissipation. Consequently, the relative depth is
\begin{eqnarray}
\frac{\Delta\eta}{\eta_{c}}&\equiv&\frac{\eta_{c}-\eta_{\textrm{min}}}{\eta_{c}}\nonumber\\
&=&\frac{\Gamma(\Gamma+\Lambda)}{\Gamma(\Gamma+\Lambda)+[2\sqrt{2}v-(2\Gamma+\Lambda)](2\Gamma+\Lambda)}.
\end{eqnarray}
Figures~\ref{fig_2site_nodisorder}(b-d) illustrate the dependence of dip position and relative depth on two different ratios $\Lambda/\Gamma$ and $(2\Gamma+\Lambda)/v$. In the same figures, we include the transfer efficiency with a nonzero local dephasing $\gamma/v=0.2$ (blue lines) for comparison. Incorporating local dephasing $\gamma$ shifts the minimum to the left and could make the decreasing part of the curve invisible. At the same time, it also makes the contrast between the dip and saturation more evident.

\begin{figure}[t]
\begin{center}
\includegraphics[width=.9\linewidth]{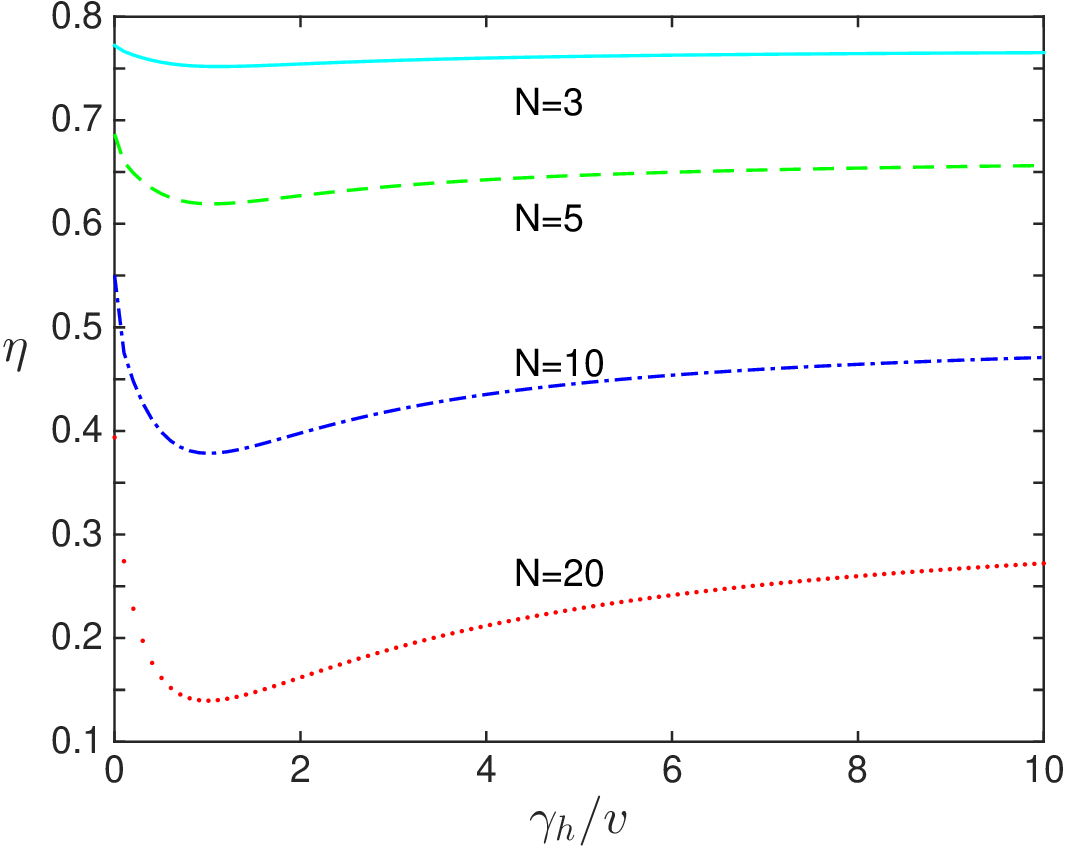}
\caption{Energy transfer efficiency as a function of $\gamma_{h}/v$ for a uniform ($\delta=0$) chain of various length $N$. Local dephasing $\gamma$ is taken to be zero. For all the plots, $\Lambda/\Gamma=10$ and $\Gamma/v=0.02$.}
\label{fig_multisite}
\end{center}
\end{figure}

For longer chains, analytic results similar to Eq.~(\ref{eq_eta_2site}) can in principle be obtained. However, the tedious closed form results are not illustrative. Instead, we resort to the numerical evolution of the master equation. Figure~\ref{fig_multisite} presents the simulation for chains of $N=3$, $5$, $10$, and $20$ sites respectively. As $N$ increases, the efficiency decreases while the relative depth of the minimum increases.

\section{Disordered system}\label{sec_disordered}
\begin{figure}[t]
\begin{center}
\includegraphics[width=.9\linewidth]{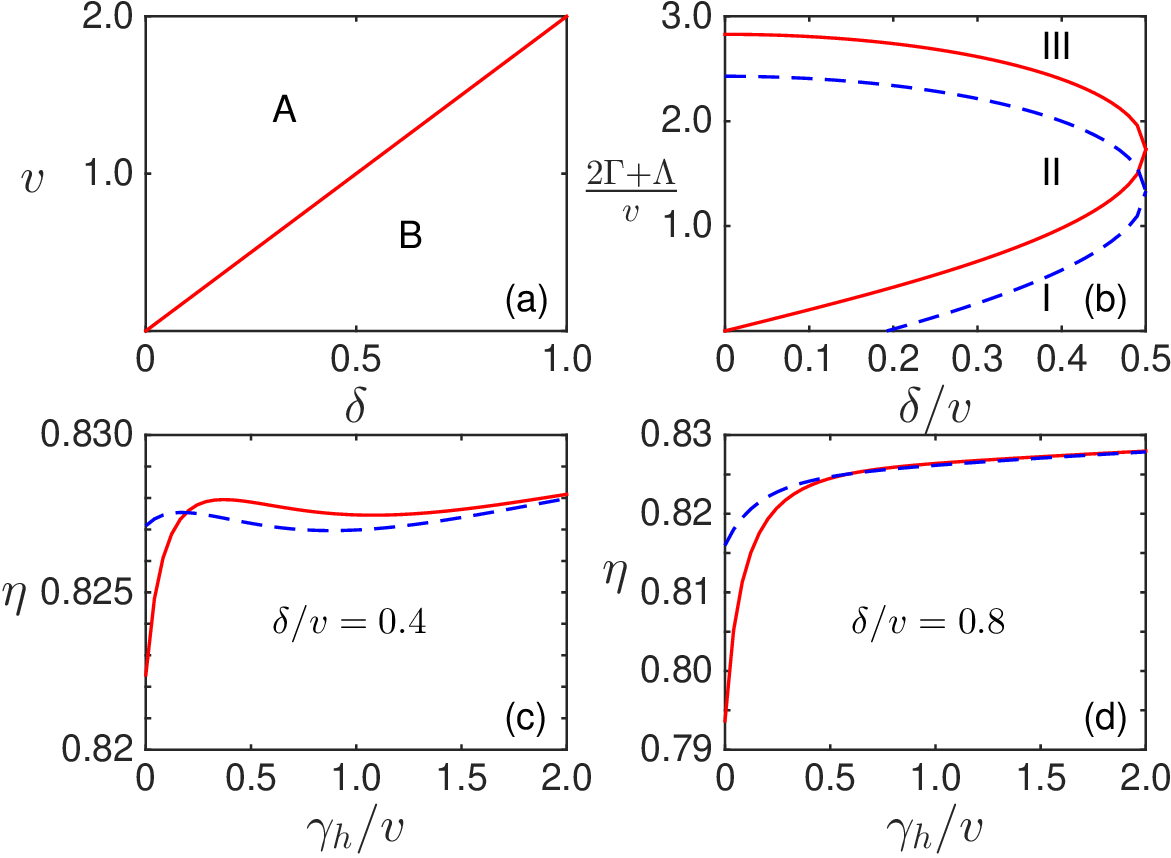}
\caption{(Color online) (a) and (b) Parameter space partition for different types of transfer efficiency as a function of the ratio of incoherent vs coherent hopping for a generic two-site ($N=2$) chain. In the lower right region of $v-\delta$ space denoted by B, efficiency monotonically increases with $\gamma_{h}/v$ as seen in (d). In contrast, in the upper left region A, for suitable combination of other incoherent rates [in (b)], efficiency may show local extrema by varying $\gamma_{h}/v$. In (b), dissipation rate $2\Gamma+\Lambda$ and disorder $\delta$ space is divided by $\mathcal{D}_{\pm}$ into region I, II, and III where efficiency displays two, one, and no extrema. (c) and (d) Efficiency as a function of $\gamma_{h}/v$ for two typical parameter sets. In both subplots, $\Lambda/\Gamma=10.0$, $\Gamma/v=0.02$, red (solid) lines for $\gamma=0$ and blue (broken) lines for $\gamma/v=0.2$.}
\label{fig_2site_phase}
\end{center}
\end{figure}

Now we turn to systems whereby disorder $\delta$ plays a role. It is seen from Eq.~(\ref{eq_eta_2site}) that $\eta$ is a monotonic increasing function of the combined parameter
\begin{equation}\label{eq_hopping}
f=\frac{4v^{2}}{\Delta}+\gamma_{h}.
\end{equation}
Simple algebra shows that depending on the ratio $\delta/v$, the efficiency as a function of $\gamma_{h}/v$ is basically categorized to two different types: (1) when $\delta/v>1/2$ [region B in Fig.~\ref{fig_2site_phase}(a)], the efficiency curve does not have local minimum or maximum; (2) when $\delta/v<1/2$ [region A in Fig.~\ref{fig_2site_phase}(a)], there may be a local maximum potentially lurking in the curve in addition to the aforementioned local minimum, although the existence of local extrema will depend on other parameters.

Figure~\ref{fig_2site_phase} shows the parameter space partition for different types of transfer efficiency as a function of $\gamma_{h}/v$ and the characterization of the extrema for a two-site ($N=2$) chain. In Fig.~\ref{fig_2site_phase}(a), the lower right region of the $v$-$\delta$ space is the territory where efficiency displays no local maximum. Figure~\ref{fig_2site_phase}(b) further delineates three regions I, II, and III when disorder is not large compared to the coherent hopping ($\delta/v<0.5$), where efficiency as a function of $\gamma_{h}/v$ shows both local extrema, only local minimum, and no local extrema, respectively. As before, the blue broken line illustrates that the nonzero local dephasing shrinks the region where extrema actually show up. In Fig.~\ref{fig_2site_phase}(c) we deliberately choose $\delta/v=0.4$ and $\gamma=0$, which falls into the region I of Fig.~\ref{fig_2site_phase}(b), and plot the efficiency as a function of $\gamma_{h}/v$. As expected, both extrema show up. Figure~\ref{fig_2site_phase}(d) presents another efficiency curve for the parameter set in the region B of Fig.~\ref{fig_2site_phase}(a).

\begin{figure}[t]
\begin{center}
\includegraphics[width=.9\linewidth]{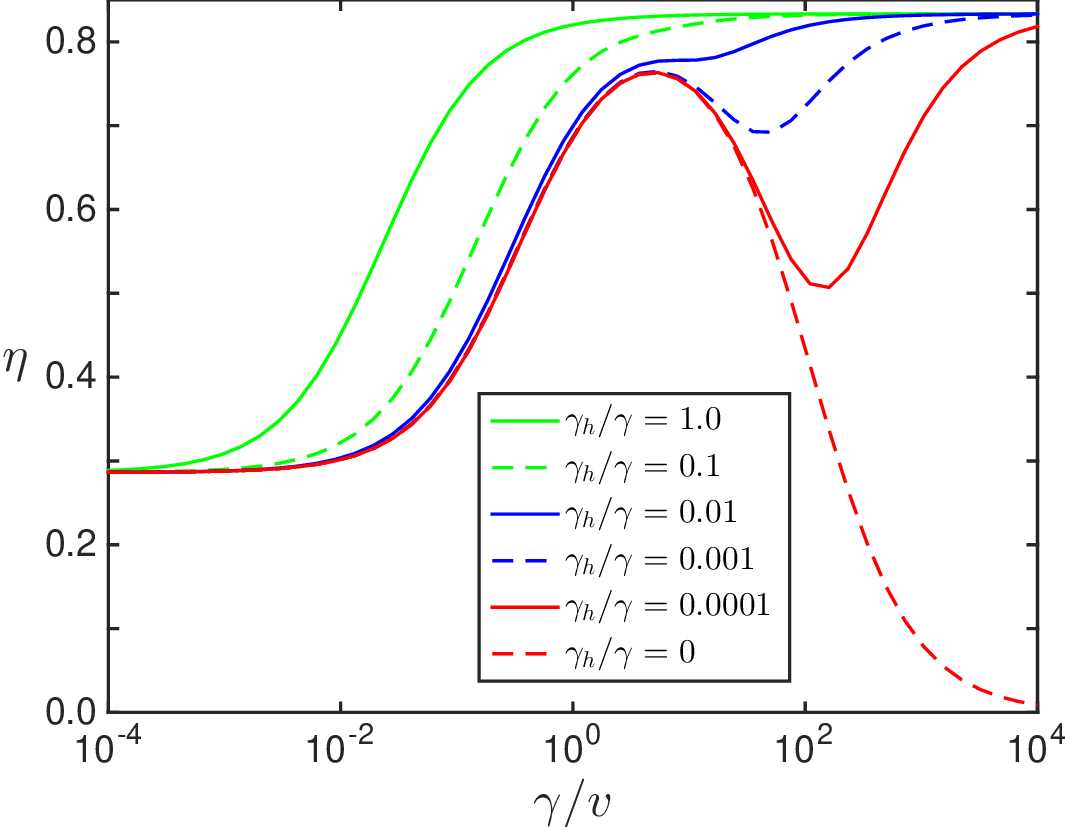}
\caption{(Color online) Energy transfer efficiency as a function of local dephasing rate as from analytical formula Eq.~(\ref{eq_eta_2site}) for various ratio $\xi$ of hopping fluctuation to local dephasing rate in disordered $\delta/v=5.0$ two-site system, where $\Lambda/\Gamma=10.0$ and $\Gamma/v=0.02$.}
\label{fig_2site_ete}
\end{center}
\end{figure}

The incoherent hopping $\gamma_{h}$ has been neglected in previous studies, on the assumption that it is much smaller than the local dephasing $\gamma$. We find that a small ratio $\gamma_{h}/\gamma$ does lead to a visible competition between the coherent and incoherent hopping mechanisms. Figure~\ref{fig_2site_ete} plots the EET efficiency as a function of $\gamma/v$ for various $\gamma_{h}/\gamma$. The dips are the turning points where ``dressed" coherent dynamics totally gives way to incoherent hopping dynamics, such that the system can effectively be described by a classical master equation instead of the Lindblad master equation.

From Eq.~(\ref{eq_hopping}), we infer that when the two additive terms are equal, the system is balanced by two separate influences, dressed coherent dynamics and purely incoherent dynamics. Therefore by solving $4v^{2}/\Delta=\gamma_{h}$, we obtain the critical incoherent hopping $\gamma_{h}^{c}$. We find that there exists a practical parameter region where $\gamma_{h}^{c}$ reaches the smallest scale as common belief states. The combined parameter $f$ can thus be seen as an effective hopping rate due to the competition among various coherent and incoherent processes within the one-excitation manifold.

\begin{figure}[t]
\begin{center}
\includegraphics[width=.9\linewidth]{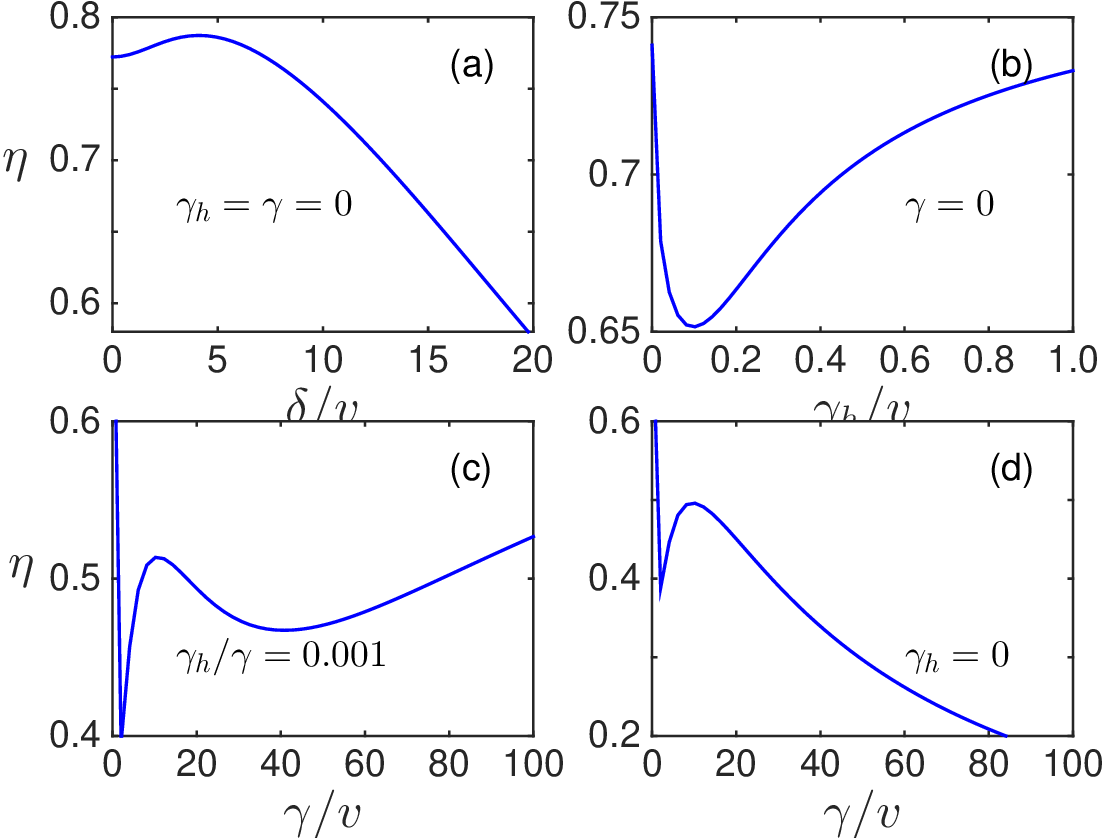}
\caption{Energy transfer efficiency in a three-site Lambda-type chain as a function of (a) ratio of disorder to cohorent hopping for $\gamma_{h}=\gamma=0$; (b) ratio of incoherent hopping to coherent hopping for $\gamma=0$; (c) ratio of local dephasing to coherent hopping for $\gamma_{h}=0.001\gamma$; and (d) ratio of local dephasing to coherent hopping for $\gamma_{h}=0$. For (b)-(d), $\Lambda/\Gamma=10$, $\Gamma/v=0.02$, and $\delta/v=10$.}
\label{fig_3site_ete_disordered}
\end{center}
\end{figure}

Figure \ref{fig_3site_ete_disordered} displays results for a disordered Lambda-type three-site chain. We see that a nonzero saturated transfer efficiency rate $\eta_{c}$, while in the absence of incoherent hopping, strong enough local dephasing leads to zero efficiency. 

\section{Experimental Realization}\label{sec_expt}
Throughout this work, we have scaled system disorder and all incoherent rates with respect to the coherent interaction between neighboring spins. To make contact with experimentally realizable conditions, we need to know the typical coherent interaction strength. In a recent work by Z. Meir \etal~\cite{ozeri}, coherent coupling between distant Sr$^{+}$ ions was attributed to the far-field resonant dipole-dipole interaction. They performed a direct spectroscopic measurement of the cooperative Lamb shift at the $5S_{1/2}\leftrightarrow5P_{1/2}$ optical dipole transition frequency. The shift can be expressed as the expectation value of the interaction in the excited state. For two ions, the cooperative shift was predicted to be 130 kHz at a distance of about 5 $\mu$m. However, in other ion-trap experiments, depending on the specific ion species and different trapping techniques, the coherent dipole-dipole interaction can be dramatically different. Recently, two ion-rap experiments aimed to produce tunable long-range coupling with $^{171}$Yb$^{+}$~\cite{monroe} and $^{40}$Ca$^{+}$~\cite{zoller}. The former experiment utilized the $^{2}S_{1/2}|F=0,m_{F}=0\rangle$ and $^{2}S_{1/2}|F=1,m_{F}=0\rangle$ hyperfine `clock' states in a magnetic dipole transition, while the latter experiment used the $|S_{1/2},m=+1/2\rangle$ and $|D_{5/2},m=+5/2\rangle$ in an electric quadrupole transition. By adjusting a  combination of trap and laser parameters, both groups can vary the interaction range in addition to actually providing nearest-neighbor interaction, on the order of 0.1 to 1 kHz. In Ref.~\cite{zoller}, it is pointed out that electric field noise leads to imperfect conservation of excitation number, while laser-frequency and magnetic-field fluctuations give rise to dephasing during the dynamics in the linear Paul trap.

\section{Conclusion}\label{sec_conclusion}
We have studied excitation transfer in a one-dimensional chain subject to both local dephasing noise and incoherent hoppping. An analytical form of the energy transfer efficiency is given for a two-site chain, which predicts rich behavior due to the competition between different processes. The main features also appear in longer chain. The phenomena predicted can be tested in quantum many-body system of trapped atomic ions recently reported to exhibit  cooperative Lamb shift~\cite{ozeri} or tunable for long-range interactions~\cite{monroe,zoller}. 

The quantum walks model we have based this work on adopts the Born-Markov approximation and deals with the dynamics to second order in the system-bath coupling. There have been several groups ~\cite{huelga, eisfeld, eisfeld11, eisfeld11njp} trying to develop fully non-perturbative and non-Markovian approaches for the open system quantum dynamics. We expect the non-local fluctuation introduced in this work to be an indispensable part of the more generic theory.

This work was supported by the NSF through a grant to ITAMP. M.Z. is partially supported by National Natural Science Foundation of China under Grant No. 11174040 and also acknowledges the State Scholarship Fund of China for additional financial aid and ITAMP for generous hospitality.

%\bibliography{reference}
%

\end{document}